\documentclass[12pt]{article}
\usepackage{lineno,hyperref}
\usepackage{graphicx}
\usepackage{braket}
\usepackage[caption=false]{subfig}
\usepackage{amsmath}

\modulolinenumbers[5]

\begin{document}

\title{Application of quantum neural network model to a multivariate regression problem}

\author{Hirotoshi Hirai\thanks{e-mail: hirotoshih@mosk.tytlabs.co.jp}\\
\\
\textit{Toyota Central R\&D Labs., Inc.,}\\
\textit{41-1, Yokomichi, Nagakute, Aichi 480-1192, Japan}}

\maketitle

\begin{abstract}
Since the introduction of the quantum neural network model, it has been widely studied due to its strong expressive power and robustness to overfitting.
To date, the model has been evaluated primarily in classification tasks, but its performance in practical multivariate regression problems has not been thoroughly examined.
In this study, the Auto-MPG data set (392 valid data points, excluding missing data, on fuel efficiency for various vehicles) was used to construct QNN models and investigate the effect of the size of the training data on generalization performance.
The results indicate that QNN is particularly effective when the size of training data is small, suggesting that it is especially suitable for small-data problems such as those encountered in Materials Informatics.
\end{abstract}

\section{Introduction}
Recently, there has been a surge of interest in quantum machine learning (QML), a technique to tackle machine learning problems through the use of quantum computing and quantum information processing~\cite{biamonte2017quantum, schuld2019quantum}.
The Harrow-Hassidim-Lloyd (HHL) method~\cite{harrow2009quantum} is capable of solving the linear equation $Ax = b$ for an $N \times N$ matrix $A$ on a time scale of $O(poly(\log N))$, which is expected to be exponentially faster than the conjugate gradient method ($O(N)$), which is the most efficient classical algorithm currently available.
Quantum linear regression~\cite{schuld2016prediction, wang2017quantum} and quantum support vector machines~\cite{rebentrost2014quantum} have been suggested as QML techniques that capitalize on this.
However, these methods necessitate a great deal of quantum gates and cannot be executed on a noisy intermediate scale quantum (NISQ) device, thus we must wait for the advent of a fault-tolerant quantum computer (FTQC).
On the other hand, a quantum neural network (QNN) method~\cite{abbas2021power}, also known as quantum circuit learning (QCL)~\cite{mitarai2018quantum}, has been proposed as an algorithm for a NISQ device.
It is a quantum-classical hybrid algorithm based on the variational quantum algorithm.
The QNN attempts to reduce the discrepancy between the output of the quantum circuit and the labeled data by adjusting the circuit parameters to their optimal values.
The advantage of QNN is that it is able to utilize high-dimensional quantum states as trial functions, which are difficult to generate on a classical computer~\cite{mitarai2018quantum}.
Another advantage is that the unitarity of quantum circuits serves as a regularization to prevent overfitting~\cite{mitarai2018quantum}.
In a conventional neural network model, a regularization term is added to the cost function to limit the norm of the learning parameter and reduces the expressibility of the model to prevent overfitting.
On the other hand, in a QNN model, the norm of parameters is automatically limited to 1 due to unitarity, so it can be said that the regularization function is inherently provided.
The exploration of QNN models is a relatively new field, and previous studies on the development of regression models have been restricted to basic single-dimensional functions~\cite{mitarai2018quantum, chen2021expressibility, kobayashi2022overfitting}.
To clarify the performance of QNN models in practical multivariate regression analysis problems, we constructed QNN models on the Auto-MPG data set (mileage per gallon data for various vehicles, 392 valid data excluding missing data)~\cite{misc_auto_mpg_9} and investigated the effect of the training data size on generalization performance.

\section{Method}
\subsection{Auto-MPG dataset}
In order to evaluate the effectiveness of QNN models in multivariate regression, the Auto-Mpg data set~\cite{misc_auto_mpg_9} from the UCI Machine Learning Repository was used.
This data set consists of 392 instances related to the city-cycle fuel consumption in miles per gallon (MPG).
It includes seven attributes that can be used to predict the MPG: cylinders (number of cylinders), displacement (displacement), horsepower (horsepower), weight (vehicle weight), acceleration (time required for acceleration 0-60 mph), model year (year of manufacture), and origin (country of production).
Although some attributes are discrete-valued, they were not encoded using one-hot encoding (encoded as separate binary variables), but they were standardized and normalized (-1,1) in the same way as continuous-valued attributes and used as input values to the model.
We conducted three experiments with different sizes of the training data set (1/5$\times$392=78, 2/5$\times$392=156 and 4/5$\times$392=312, and the remaining data were used as test data) to study the effect of size on generalization performance~\cite{yang2017prediction}.

\subsection{QNN models}
\begin{figure}[ht]
\centering
\includegraphics[width=14cm]{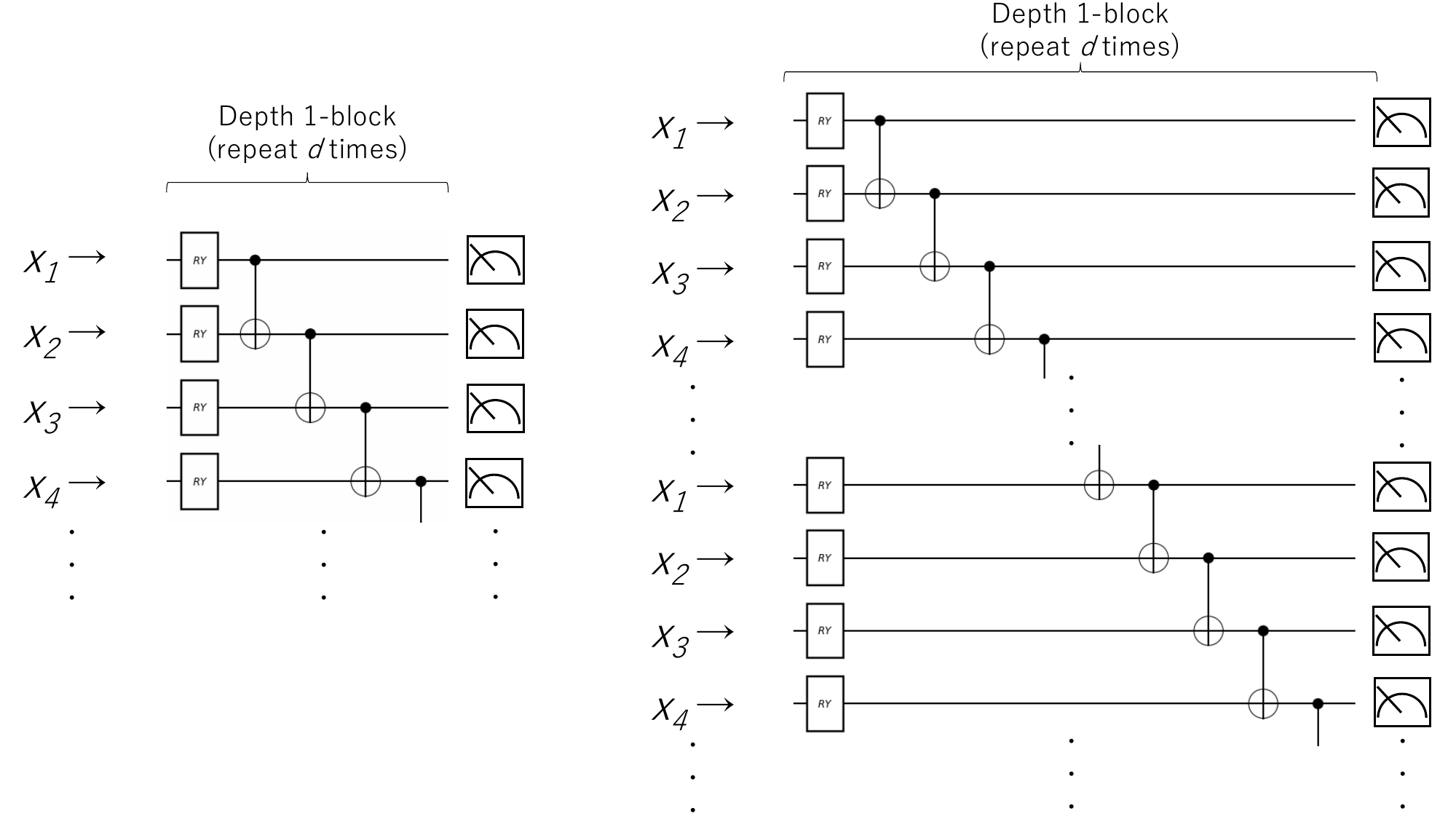}
\caption{Quantum circuits (ansatz) used for QNN models, the left side: 7-qubit model ($w=1$), the right side: 14-qubit model ($w=2$).}
\label{fig_circuit}
\end{figure}

The QNN architecture is composed of three components: an encoder which transforms classical data into a quantum state, an ansatz which is a quantum circuit with learning parameters, and a decoder which converts the quantum state into an output value.
Here, the Ry rotation gate (the Ry gate acting on each qubit initialized to $\ket{0}$) was used as the encoder.
The rotation angle was set to $\theta$=arctan($x$)+$\pi/2$ for the scaled attribute $x$.
The arctangent allows the scaled attribute to be uniquely converted to a rotation angle even if the value is outside the range (-1,1) when the scaler is used for the test data.
In this study, we constructed the 7-qubit model with each attribute encoded in one qubit (circuit width $w=1$) and the 14-qubit model with each attribute encoded in two qubits ($w=2$), as shown in Fig.~\ref{fig_circuit}.
For an ansatz, we used the circuit with Ry rotation gates and CNOT gates in linear configuration (see Fig.~\ref{fig_circuit}).
The learning parameters are the rotation angles of the Ry rotation gates.
We built the model by connecting $d$ of the depth 1-blocks ($d$ is referred to as the depth of the circuit).
The sum of the Z-axis projections of each qubit, $\sum_i \sigma^i_z$, was used as a decoder.

The QNN models were implemented with Pytket~\cite{sivarajah2020t}, a Python module for quantum computing, and the quantum circuit calculations were performed using state vector calculations with the Qulacs~\cite{suzuki2021qulacs} backend, a quantum computing emulator.
The mean squared error (MSE) between the teacher data and the predictions was used as a cost function.
The Powell method was used to optimize the learning parameters.

\subsection{Classical NN models}
A conventional neural network (NN) model was also constructed for comparison.
We employed the NN model consisting of an input layer with 7 nodes, two hidden layers with 100 nodes each, and an output layer with a single node.
The ReLU was used for the activation functions.
PyTorch~\cite{imambi2021pytorch} was used to build and train the NN model.
The Adam optimizer~\cite{kingma2014adam}, an extended version of stochastic gradient descent, was used with a learning rate of 0.02 over 10000 epochs. 
L2 regularization was applied to prevent overfitting.
The training data were divided into two parts, with 80\% used for training and the remaining 20\% used for validating the L2 regularization weight parameter.

\section{Results and discussion}
\begin{figure}[ht]
\centering
\includegraphics[width=14cm]{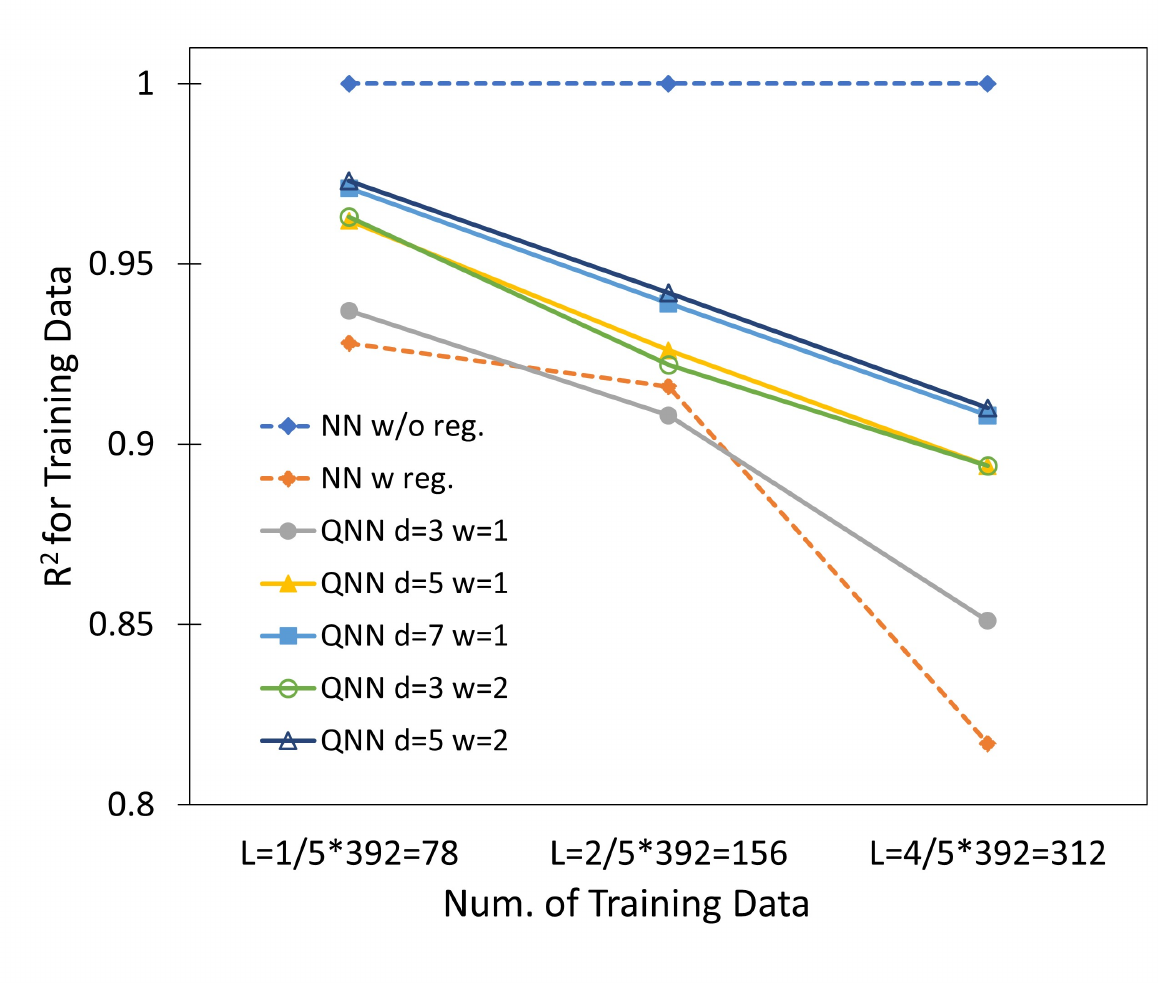}
\caption{$R^2$ values for the training data.}
\label{fig_train}
\end{figure}

The $R^2$ (coefficient of determination) for the training data (expressivility) for each QNN model are shown in Fig.~\ref{fig_train}.
Figure~\ref{fig_train} also shows the results of the classical NN model) for comparison.
The QNN model with a deeper (larger $d$) or wider (larger $w$) circuit has stronger expressivility.
The $R^2$ of QNN models decreases as the size of the training data increases.
This is the same behavior as the classical NN model with regularization.
On the other hand, the classical NN model without regularization shows a perfect fit ($R^2=1$) at any data size.
These results indicate that the automatic regularization worked in the QNN models.

\begin{figure}[ht]
\centering
\includegraphics[width=14cm]{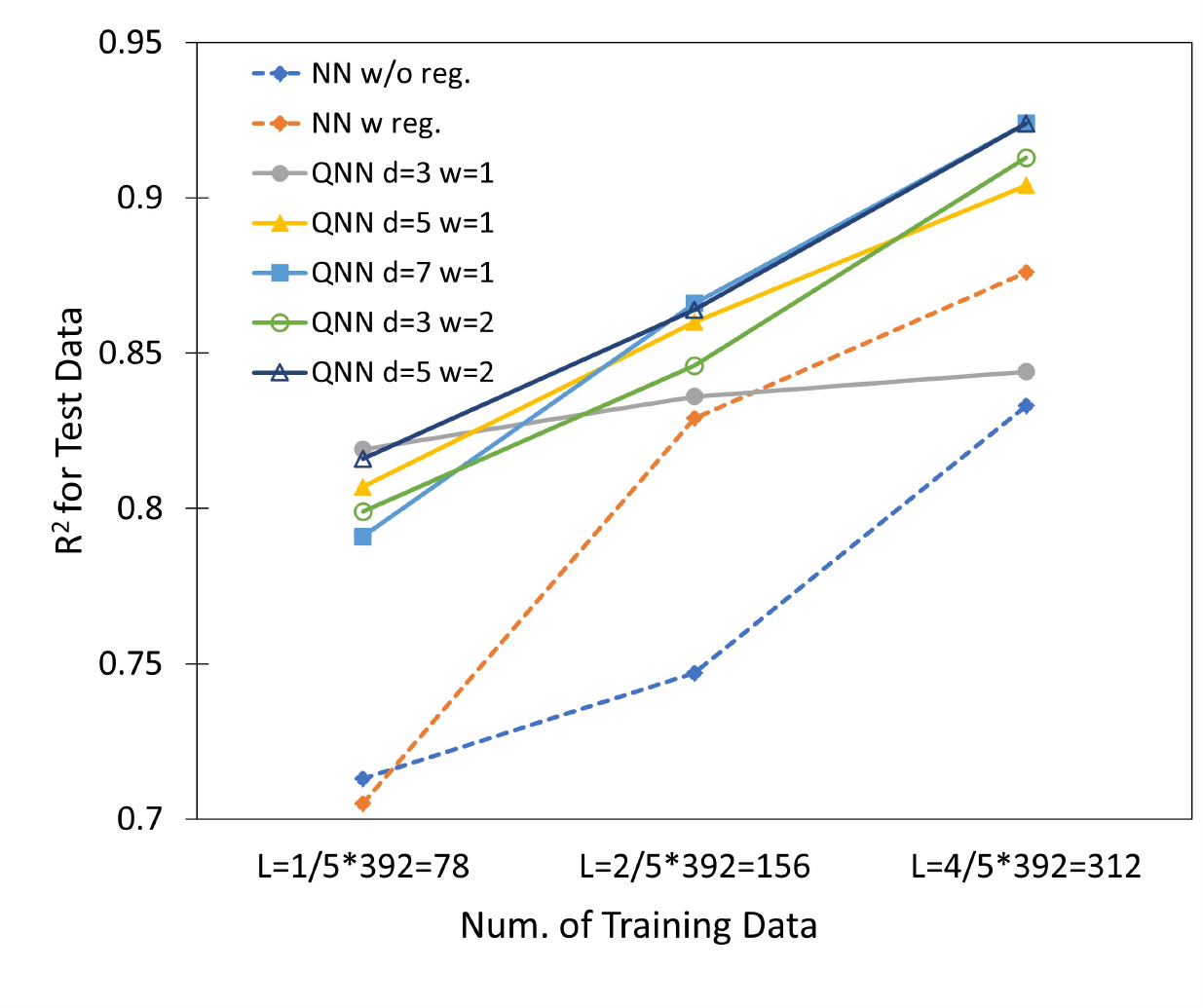}
\caption{$R^2$ values for the test data.}
\label{fig_test}
\end{figure}

The $R^2$ for the test data (generalization performance) for each QNN model and the classical NN models are shown in Fig.~\ref{fig_test}.
QNNs have been shown to have superior generalization performance compared to classical NN models when the amount of training data is limited.
As the size of the training data increases, the gap between the QNN and the classical NN models narrows, and the smallest QNN model ($d = 3$, $w = 1$) performs worse than the classical NN model when the largest data size.
The smallest QNN model has limited expressibility due to its limited number of learning parameters. Conversely, a QNN model with a deeper or wider circuit has enough expressibility, and its generalization performance is superior to that of classical NN models.
This distinction is particularly evident when the amount of training data is limited and the QNN model is considered to be especially effective for small datasets.
This advantage has also been confirmed in classification problems~\cite{caro2022generalization}.
Thus, QNNs are considered to be particularly promising for Materials Informatics (MI) problems with a small number of data~\cite{butler2018machine,xu2023small}.
The use of QNNs offers a great advantage in that they do not need to be regularized and can still achieve excellent generalization performance without the need for hyperparameter tuning.
Circuit width and depth may be adjustable parameters, but it appears that they should be set to the same level as the number of attributes.
Since the problems handled by typical MI do not require a large number of attribute variables and can be adequately computed by emulation on a classical computer without using an actual quantum computer, it may be used as a quantum-inspired algorithm.
As a challenge, QNNs have been thought to require more time to train than classical NN models.
However, a recent study~\cite{abbas2021power} has shown that, with the use of a specific ansatz, QNNs can be more trainable than classical NNs, raising high expectations for their future progress.

\section{Conclusion}
We constructed QNN models on the Auto-MPG data set to clarify the performance of QNN models in practical multivariate regression analysis problems.
Compared to classical NN models, QNNs showed better generalization performance when the size of the training data was limited.
The results suggest that QNN is particularly effective when the data size is small, suggesting that it is especially suitable for small data problems such as those encountered in Materials Informatics.

\bibliography{ref}

\end{document}